\documentclass[pdflatex,sn-mathphys,smallcondensed]{sn-jnl}
\jyear{2021}
\usepackage{graphicx}
\usepackage{subfigure}
\theoremstyle{thmstyleone}

\usepackage{tabularx}

\theoremstyle{thmstyletwo}

\theoremstyle{thmstylethree}

\usepackage{ulem}

\begin{document}

\title[DACSR: Decoupled-Aggregated End-to-End Calibrated Sequential Recommendation]{DACSR: Decoupled-Aggregated End-to-End Calibrated Sequential Recommendation}

\author[1]{\fnm{Jiayi} \sur{Chen}}\email{jychen@ica.stc.sh.cn}
\author*[1,2]{\fnm{Wen} \sur{Wu}}\email{wwu@cc.ecnu.edu.cn}
\author[1]{\fnm{Liye} \sur{Shi}}\email{lyshi@ica.stc.sh.cn}
\author[1]{\fnm{Yu} \sur{Ji}}\email{52205901009@stu.ecnu.edu.cn}
\author[3]{\fnm{Wenxin} \sur{Hu}}\email{wxhu@cc.ecnu.edu.cn}
\author[2]{\fnm{Xi} \sur{Chen}}\email{xchen@psy.ecnu.edu.cn}
\author[4]{\fnm{Wei} \sur{Zheng}}\email{wzheng@admin.ecnu.edu.cn}
\author[1]{\fnm{Liang} \sur{He}}\email{lhe@cs.ecnu.edu.cn}

\affil[1]{\orgdiv{School of Computer Science and Technology}, \orgname{East China Normal University}, \orgaddress{\postcode{200062}, \state{Shanghai}, \country{China}}}

\affil[2]{\orgdiv{Shanghai Key Laboratory of Mental Health and Psychological Crisis Intervention, School of Psychology and Cognitive Science}, \orgname{East China Normal University}, \orgaddress{\postcode{200062}, \state{Shanghai}, \country{China}}}

\affil[3]{\orgdiv{School of Data Science and Engineering}, \orgname{East China Normal University}, \orgaddress{\postcode{200062}, \state{Shanghai}, \country{China}}}

\affil[4]{\orgdiv{Information Technology Services}, \orgname{East China Normal University}, \orgaddress{\postcode{200062}, \state{Shanghai}, \country{China}}}

\abstract{Sequential recommendations have made great strides in accurately predicting the future behavior of users. However, seeking accuracy alone may bring side effects such as unfair and overspecialized recommendation results. In this work, we focus on the calibrated recommendations for sequential recommendation, which is connected to both fairness and diversity. On the one hand, it aims to provide fairer recommendations whose preference distributions are consistent with users' historical behaviors. On the other hand, it can improve the diversity of recommendations to a certain degree. But existing methods for calibration have mainly relied on the post-processing on the candidate lists, which require more computation time in generating recommendations. In addition, they fail to establish the relationship between accuracy and calibration, leading to the limitation of accuracy. To handle these problems, we propose an end-to-end framework to provide both accurate and calibrated recommendations for sequential recommendation. We design an objective function to calibrate the interests between recommendation lists and historical behaviors. We also provide distribution modification approaches to improve the diversity and mitigate the effect of imbalanced interests. In addition, we design a decoupled-aggregated model to improve the recommendation. The framework assigns two objectives to two individual sequence encoders, and aggregates the outputs by extracting useful information. Experiments on benchmark datasets validate the effectiveness of our proposed model.}
\keywords{Sequential Recommendation, Calibrated Recommendation, Fairness, Diversity}

\maketitle

\section{Introduction}\label{sec:introduction}

Recommender systems aim to help users find their interests among large-scale items. In recent years, sequential recommendation has achieved great attention, which predicts users' future behaviors according to sequences of historical behaviors. Existing studies focus on modeling sequences, learning item transitions and obtaining accurate recommendations. The deep learning-based architectures, such as Recurrent Neural Networks and Graph Neural Networks, have progressed in sequential recommendation \cite{hidasi2015session, chang21seqgnn,wu2018session}. Until now, most studies have focused on obtaining high accuracy of recommendation lists. However, previous studies argue that recommendation algorithms should consider more than accuracy. For example, diversity \cite{huang2018improving, lu21fat, cen20comirec,zheng21dgcn, parapar21unifiedmetric}, coverage \cite{tailnet, ijcai19seqdiverse,sreepada21mitigatinglongtail}, unexpectedness \cite{chen19seren, xu20neuralseren,li20purs} and fairness \cite{dai21fairgnn, ge21longtermfair} are also important concepts in measuring a recommender system. 

Among these concepts, diversity requires the recommender system to generate item lists that contain more item attributes (e.g., genres of movies), and a fair recommender can provide unbiased recommendation lists for consumers or providers. From these two perspectives, we focus on the calibration of sequential recommendation, which is related to diversity and fairness. The calibrated recommendation was first proposed in \cite{steck18calibrated}. It aims to provide the recommendation list which reflects the user's historical behaviors \cite{steck18calibrated}. For example, if a user has watched 70\% action movies and 30\% comedies, a fully calibrated recommendation list should also contain action and comedy movies with this ratio. Compared to diversity, it can also provide diversified recommendation lists to a certain degree \cite{Kaya19comparison}. The difference between calibration and diversity is that calibration limits the covered item attributes which follow the user's historical behaviors. It is a type of C-fairness according to the taxonomy proposed by \cite{burke17multisidefair}, which is the fairness from the consumer's perspective. For users with similar historical interests, the calibrated recommendation model is able to provide recommendation lists with similar interest distributions. This somehow avoids bias and reflects the fairness of the recommendation system.

\begin{figure}[tbp]
    \centering
    \includegraphics[width=10cm]{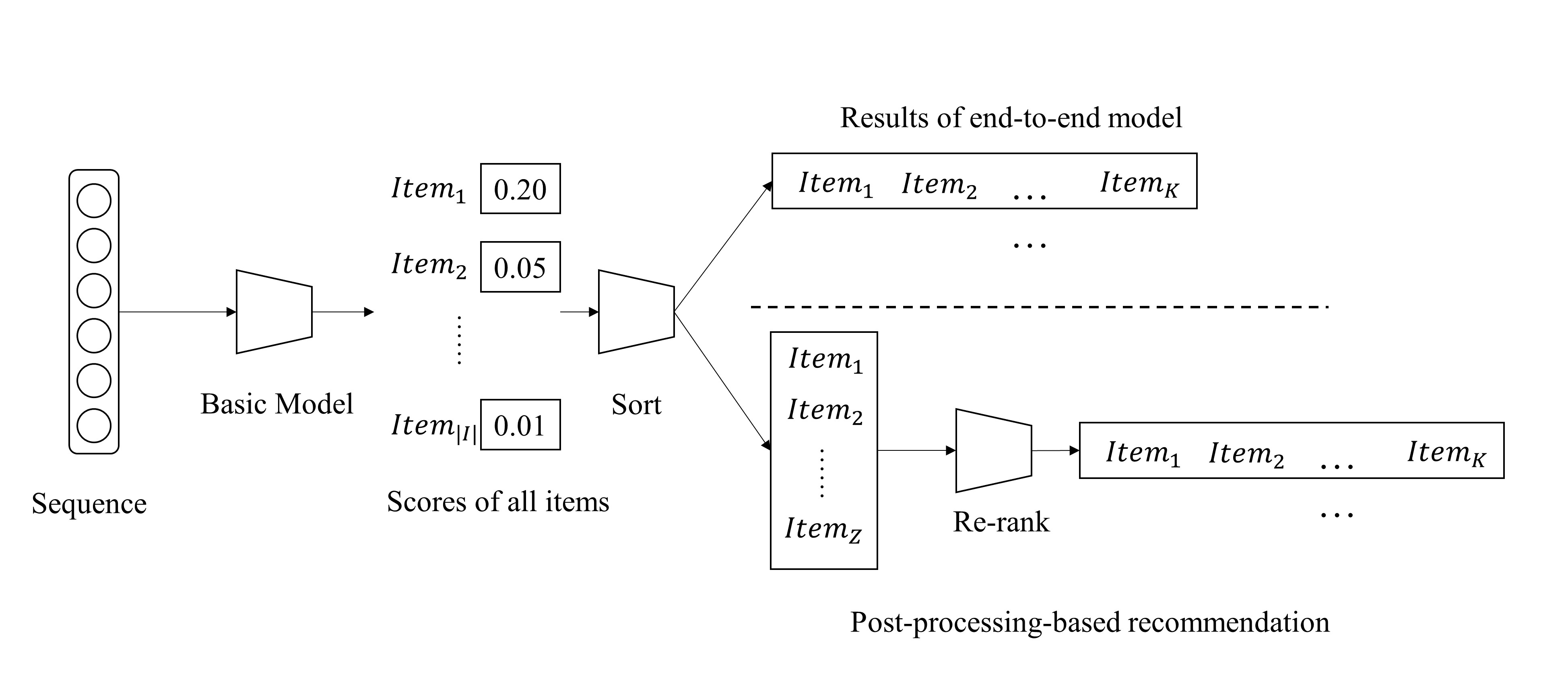}
    \caption{An illustration of end-to-end and post-processing-based recommendation}
    \label{fig:end-to-end}
\end{figure}

Existing studies in calibrated recommendations adopt a post-processing paradigm, which ranks items from a candidate list that has been generated by a basic recommendation model such as neural collaborative filtering \cite{steck18calibrated,seymen21constrain, silva21exploiting}. This is different from the end-to-end recommendation paradigm of sequential recommendation. As illustrated in Fig. \ref{fig:end-to-end}, both end-to-end and post-processing-based models require predicted scores of all items. The difference is that the end-to-end models directly select the top-K items as final recommendations, while post-processing-based models apply a re-ranking stage on the top-Z (Z $>$ K) items. For example, Kaya et al. \cite{Kaya19comparison} and Silva et al. \cite{silva21exploiting} generated calibrated recommendations based on the top-100 items provided by the basic algorithms. The advantage of post-processing-based approaches is that they can be applied to almost any recommendation model, because they only require the items and scores of the candidate list generated by the base model. However, the post-processing-based models have two limitations. On the one hand, these models do not consider the relation between accuracy and calibration. The post-processing-based models make calibrated predictions based on the trained basic models which are optimized by the accuracy objective. In this scenario, accuracy and calibration are separate objectives, so that recommendation accuracy is limited by the trained basic model and considering calibration may sacrifice the accuracy. On the other hand, these models need more response time in generating recommendation lists because they have one more extra ranking stage than the end-to-end models. During the ranking stage, the items of the final recommendation list are decided step by step, in which the model needs to compute the gains of all candidates. Though the number of candidates can be much less than the size of the item set by selecting top-Z items, the ranking stage is still time-consuming.

To handle these problems, we focus on providing accurate and calibrated recommendation lists in an end-to-end framework. We propose a \textbf{D}ecoupled-\textbf{A}ggregated \textbf{C}alibrated \textbf{S}equential \textbf{R}ecommendation framework, namely ``\textbf{DACSR}''. First, we define a loss function for calibration to allow models can be optimized by accuracy and calibration simultaneously. We combine the prediction scores of all items and the item attribute information to estimate the distribution of the recommendation list. Then we use cosine similarity to measure the consistency between distributions of the recommendation list and behavior sequence. 
We also provide distribution modification approaches to improve the diversity and mitigate the problem of amplification of main interests. Next, handle the objectives of accuracy and calibration, we propose a decoupled-aggregated framework to provide accurate and calibrated recommendations. We utilize two individual sequence encoders which only focus on accuracy and calibration, respectively. Then we concatenate the item embeddings and sequence representations, and obtain final representations of the sequence and items by extractor networks with feed-forward networks and residual connections. The scores of all items are computed by the final representations, and are optimized according to the weighted sum of accuracy and calibration loss functions.

The contributions of this paper are listed as follows:
\begin{itemize}
    \item We propose an end-to-end framework to provide accurate and calibrated recommendation lists for sequential recommendation.
    \item We design a calibration loss function for model optimization, which aligns the preference distribution of recommendations to the historical distribution. In addition, we provide distribution modification methods for diversity and imbalanced interests. 
    \item We propose a decoupled-aggregated framework which aggregates information from individual sequence encoders which are optimized by calibration and accuracy separately.
    \item Experiments on benchmark datasets show that our model can achieve accurate and calibrated recommendations, with less time consumption than post-processing-based models.
\end{itemize}
The rest of this paper is organized as follows. We first review the existing literature in Sec. \ref{sec:rel} and provide some preliminaries about sequential and calibrated recommendation in Sec. \ref{sec:rel} and \ref{sec:preliminary}. Then we introduce our model in Sec. \ref{sec:method}. We provide experimental settings in Sec. \ref{sec:experiment} and show results and analysis in Sec. \ref{sec:result}. Finally, we discuss our work in Sec. \ref{sec:discussion} and conclude our paper and indicate some future work in Sec. \ref{sec:conclusion}.

\section{Related Work}\label{sec:rel}

In this section, we provide a literature review of our work. We first introduce existing studies in sequential recommendation. Then we review the recent advances in calibrated recommendation.

\subsection{Sequential Recommendation}
Sequential recommendation relies on users' historical behavior sequences to predict their future behaviors. Existing studies focused on modeling sequences and obtained better sequence representations to achieve higher recommendation accuracy. Hidasi et al. \cite{hidasi2015session} first utilized gated recurrent units for sequential recommendations and provided a parallel training strategy. Li et al. \cite{li2017neural} further proposed an attention mechanism to capture the main purposes of the sequence. Tang et al. \cite{tang18caser} utilized convolutional neural networks to extract information from short-term sequences. With the development of self-attention, self-attentive-based models were proposed to achieve better sequence representations. For example, SASRec applied a self-attentive mechanism to learn both long-term and short-term preferences of user behavior sequences, which achieved satisfactory performance of recommendation accuracy \cite{kang18sasrec}. Xu et al. \cite{xu21longshortsa} divided sequences into subsequences and captured users' long- and short-term preferences by applying two self-attention networks. In addition, transformer-based sequence encoders were proposed, such as BERT4Rec and Transformer4Rec \cite{sun19bert4rec, moreira21transformer4rec}. In recent years, graph neural networks were also utilized for sequential recommendation \cite{wu2018session, chang21seqgnn, gcegnn}. For example, Wu et al. \cite{wu2018session} applied GGNN to learn item transitions from historical behaviors which treated sequences as graphs. In addition, multi-interest-based models were proposed that used multiple vectors to represent a sequence in order to disentangle users' diverse intentions \cite{cen20comirec, lu21fat, chen20improvingend, tan21sine}.

\subsection{Calibrated Recommendation}

Existing sequential recommendation models achieved satisfactory recommendation accuracy. From the concerns of fairness and filter bubble, we focus on the calibration of recommendation lists of sequential recommendation algorithms. In recent years, calibration was proposed, which aimed to generate recommendation lists whose preference distributions were less divergent with the users' profile \cite{steck18calibrated}. Steck \cite{steck18calibrated} also provided a post-processing greedy re-ranking model which considered both accuracy and calibration at each step of generating results. Abdollahpouri et al. \cite{himan20connection} studied the connections between popularity bias and calibration. They found that users who were affected more by popularity bias tend to achieve less calibrated recommendation lists. Kaya and Bridge \cite{Kaya19comparison} compared intent-aware algorithms and calibration algorithms. They found that the diversity-oriented intent-aware models can achieve calibrated recommendations and calibration-oriented models can obtain diversity to some extent. Seymen et al. \cite{seymen21constrain} proposed weighted total variation to measure the consistency between two distributions and a constrained optimization model to improve the ranking stage for calibration. Silva et al. \cite{silva21exploiting} proposed new metrics to evaluate calibrated recommendations and adaptive selection strategies for the trade-off weight in the post-processing algorithms. 

The calibration of recommendations is connected to two types of concepts. One is diversity, which aims to provide diversified recommendations for users. Seeking accuracy may lead to skewed recommendation lists which only focused on the main interest area of users, but users may be interested in diversified lists \cite{steck18calibrated, chen20improvingend}. The calibrated recommendation constrained the recommendations to match the user's historical preference distribution to avoid the problem. However, it is a type of limited diversity because recommendations are limited by users' historical behaviors.
Despite the limitation of interests, it is still considered as a solution of homogeneous contents\cite{Kaya19comparison,zhao21tecrec}. Calibration is also a type of fairness. Fairness in recommender systems aimed to provide unbiased results for users. From the perspective of stakeholder, it can be defined as C-fairness, P-fairness and CP-fairness, which stand for consumers, providers, and the combination, respectively \cite{burke17multisidefair}. The calibrated recommender system can be treated as one of C-fairness \cite{silva21exploiting}. The fairness is reflected by the less divergence of preference distributions between the user's profile and the recommendation list.

Despite previous advances in calibrated recommendation, it still suffers from the following problems. First, existing methods for calibration required re-ranked the candidate items generated by a basic recommendation model, which required more time in generating recommendations. In addition, the post-processing models may sacrifice accuracy to improve calibration, because they separated the process of achieving the accuracy and calibration. In our work, we would like to explore whether calibrated recommendations can be provided in an end-to-end way without post-processing. In addition, we investigate whether considering both calibration and accuracy can contribute to the performance of sequential recommendation.

\section{Preliminary}\label{sec:preliminary}
\subsection{The Sequential Recommendation Paradigm}
The sequential recommendation predicts items that the user may interact in the future based on the user's historical behavior sequence. In general, it can be decomposed into two parts, the sequence encoder and the prediction layer. The sequence encoder takes the historical behavior sequence as the input, and represents it to a vector. Formally, the procedure can be written as:
\begin{equation}
    h = f(s \mid E^I,\theta, L)
\end{equation}
where $f(\cdot)$ is the sequence encoder and $h$ is the sequence representation of sequence $s$. $E^I$ represents the item embedding matrix of all items $I$, and $\theta$ stands for the parameters of the sequence encoder. $L$ is the loss function that is used to optimize the sequence encoder $f$.

The sequence representation $h$ is further used to predict the score of all items. The prediction layer is usually a linear transformation layer:
\begin{equation}
    \hat{y} = Wh^\mathsf{T} + b 
\end{equation}
where $W$ and $b$ are $\mid I \mid \times d $ and $\mid I \mid$ dimensional learnable parameters. A common setting is that $W$ is the item embedding matrix which is used in $f$ and bias $b$ is removed:
\begin{equation}\label{eq:dot}
    \hat{y} = E^{I} h^\mathsf{T}
\end{equation}
where $E_{I}$ is the item embedding matrix. This prediction layer is widely used in existing studies \cite{li2017neural, wu2018session, kang18sasrec, xu21longshortsa, zhang21attenhybridrnn}, and we also follow this setting in our work. In our work, we select SASRec \cite{kang18sasrec} as the basic sequence encoder, because our goal is not to investigate modeling sequences and the SASRec model has achieved satisfactory performances in existing studies.

\subsection{Preference Distributions}
The calibrated recommendations aim to provide unbiased results, where preferences reflected from historical behaviors and recommendation lists are consistent. We use symbols $p(s)$ and $q(s)$ which stand for the preference distributions from historical behaviors and recommendation lists, respectively. We follow previous work which applied item attributes to define preference distributions \cite{steck18calibrated}, which are introduced below in detail.
\begin{itemize}
    \item $p(s)$ is the preference distribution from the sequence $s$. For each attribute $g$, the distribution value is computed as:
    \begin{equation}\label{eq:ps}
        p(g \mid s) = \frac{\sum_{x \in s} p(g \mid x)}{\mid s \mid}
    \end{equation}
    where $p(g \mid x)$ is the indicator function of item $x$ and attribute $g$, which satisfies $\sum_{g \in G} p(g \mid x) = 1$. If item $x$ does not contain attribute $g$, the value of $p(g \mid x)$ is 0. If the item contains two attributes, the value of $p(g \mid x)$ equals to 0.5 for each attribute $g$. Finally, the preference distribution can be represented as a $G$-dimensional vector $\{p(g=1 \mid x), p(g=2 \mid x),..., p(g=G \mid x)\}$, where $\mid G \mid$ is the total amount of item attributes.
    \item $q(s)$ is the preference distribution of the recommendation list. For each attribute $g$, the distribution value is computed as:
    \begin{equation}\label{eq:qs}
        q(g \mid RL_s) = \frac{\sum_{x \in RL} p(g \mid x)}{K}
    \end{equation}
    where $RL_s$ is the recommendation list of the sequence $s$, and $K$ is the size of $RL_s$. Similar to $p(s)$, the distribution $q(s)$ can also be represented as a $G$-dimensional vector $\{q(g=1 \mid x), q(g=2 \mid x),..., q(g=G \mid x)\}$.
\end{itemize}

\section{Methodology}\label{sec:method}

In this section, we introduce our DACSR model in detail. We first introduce our proposed calibration loss function for end-to-end sequential recommendation. Then we introduce the Decoupled-Aggregated architecture in detail.

\subsection{The Calibration Loss Function}\label{sec:loss}

\textbf{Loss Function for Calibration} To generate a calibrated recommendation list, we design the loss function for the model training. The calibration measures the consistency between the recommendation list and the historical sequence. The distribution of historical sequence $p(s)$ can be computed as Eq. \ref{eq:ps}. For the recommendation list, we estimate its preference distribution $\hat{q}(s)$ as follows:
\begin{align}
    \hat{q}(g \mid s) &= \sum_{i}^{\|I\|}\hat{y}_i \cdot p(g \mid i)\\
    \hat{y}_i &= softmax(\hat{y}_i / \tau)
\end{align}
where $\hat{y}_i$ is the score of the item $i$ predicted by the model, and it is further processed by a softmax function. If an item has a higher prediction score, it contributes more to $\hat{q}(g \mid s)$. The softmax function also amplifies the difference in scores. Items with high scores will still be given higher weights, while weights of other items are close to 0. $\tau$ ($\tau > 0$) is the temperature parameter of softmax function. If $\tau < 1$, the score distribution becomes sharper and items with higher scores get more emphasis. In contrast, an extremely large value of $\tau$ will make the score distribution more uniform.  

After estimating $\hat{q}(g \mid s) (g \in G)$, we define the loss function of calibration as:
\begin{equation}
    L_{Calib}(\hat{y}) = 1 - cos(\hat{q}(s), p(s))
\end{equation}
where $\hat{q}(s)$ is the estimated distribution vector $\{\hat{q}(g=1 \mid s),..., \hat{q}(g = G \mid s)\}$ and $cos(v_1, v_2)$ is the cosine similarity between two vectors. If the two distributions are more consistent, the value of $L_{Calib}$ will be lower.  

\textbf{Distribution Modification} Although calibration is related to diversity, a calibrated recommendation list is not always a diversified list. For example, a user who focuses on a few types of items will receive less diversified recommendations when calibration is considered. Therefore, if we want a diversified recommendation list to a certain degree, we can modify the distribution as follows:
\begin{equation}\label{eq:div}
    p_d(s) = softmax(p(s)/\tau_{div})
\end{equation}
In this equation, the historical distribution $p(s)$ is normalized by a softmax function. For item attribute $g$ that the user did not interacted (i.e., $p(g \mid s)=0$), it will obtain a positive value. Therefore, all attributes are considered. The parameter $\tau_{div}$ is also used to control the distribution, similar to $\tau$.

Meanwhile, for users who have homogeneous interests, their main interests may be amplified under the end-to-end framework. This is similar to the imbalanced classification tasks which tend to predict the major labels. To this end, we propose a mask-based modification method:
\begin{equation}\label{eq:maskdiv}
    p_m(s) = softmax(mask(p(s)) / \tau_{div})
\end{equation}
where the $mask(p(s))$ give all attributes whose $p(g \mid s)=0$ an extremely little negative value (e.g., $-10^{10}$). Therefore, scores of these attributes in $p_m(s)$ will be 0. In this equation, $\tau_{div}$ can be larger than 1 so that the distribution becomes more uniform, and the scores of the main interest and other interests are more close. The difference from Eq. \ref{eq:div} is that the scope of interests are still limited in those the user have interacted with, while the $p_d(s)$ distribution can explore new interests for the user.

\textbf{Loss Function for Accuracy and Calibration} To obtain recommendation lists with accuracy and calibration, an intuitive way is directly optimizing the sequential recommendation model with a weighted sum of loss function:
\begin{equation}
    L_w = (1 - \lambda) \times L_{Acc}(y, \hat{y}) + \lambda \times L_{Calib}(\hat{y})
\end{equation}
where $\lambda \in [0, 1]$ is the trade-off factor between accuracy and calibration. A higher value of $\lambda$ means more consideration on calibration. $L_{Acc}$ is the accuracy-based loss function. In our work, we choose the cross-entropy loss function:
\begin{equation}
    L_{Acc}(y, \hat{y}) = \sum_{i=0}^{\mid I \mid} y_i \log (\hat{y}_i)
\end{equation}
where $y$ and $\hat{y}$ are the vectors of the ground-truth and predicted scores of all items, respectively. The $y$ is an one-hot vector where $y_i=1$ means item $i$ is the next item of the sequence, and $0$ otherwise. 

\begin{figure}[tbp]
    \centering
    \includegraphics[width=10cm]{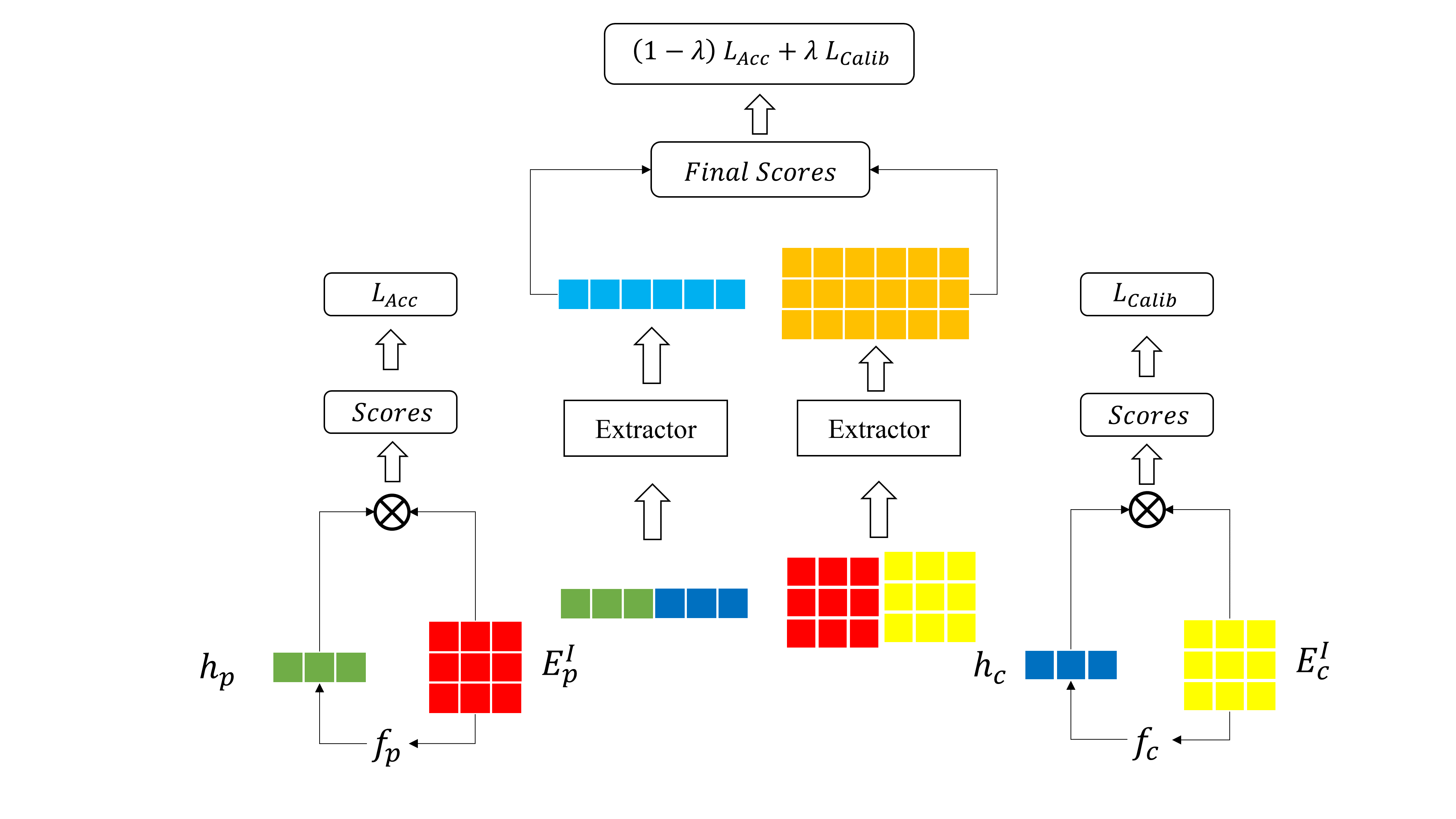}
    \caption{The architecture of our DACSR model}
    \label{fig:model}
\end{figure}

\subsection{The Decoupled-Aggregated Framework}
Directly optimizing a sequence encoder by the weighted loss function may lead to a seesaw problem \cite{ma18mmoe, tang20ple}. For example, the performance of calibration increases by sacrificing the recommendation accuracy. This is because optimizing by two objectives based on shared parameters limits the ability of the model to obtain better representations of sequences and items. Therefore, we propose our Decoupled-Aggregated framework, which includes two basic sequence encoders, as shown in Fig. \ref{fig:model}. The two sequence encoders are optimized separately. One is to accurately predict the next behavior, while the other one is to provide a fully calibrated recommendation list. Formally, this can be represented as:
\begin{align}
h_p &= f_p(s \mid E_p^{I}, \theta_p, L_{Acc})\\
h_c &= f_c(s \mid E_c^{I}, \theta_c, L_{Calib})
\end{align}
where $f_p$ and $f_c$ are two different sequence encoders with different parameters and objective functions. $E_p^{I}$ and $E_c^{I}$ are item embedding matrices of the two sequence encoders. $\theta_p$ and $\theta_c$ are their parameters. Note that the two encoders $f_p$ and $f_c$ do not share the same parameters and item embedding matrices, and are optimized by their unique loss functions $L_{Acc}$ and $L_{Calib}$, respectively.

Next, we use the sequence representations and item embeddings of the two encoders to provide calibrated recommendation list. A direct way is to concatenate the representations of two sequence encoders:
\begin{align}
    h_a = [h_p, h_c]\\
    E_a^I = [E_p^I, E_c^I]
\end{align}
Based on this, we use the concatenated vectors as input to two extractor nets:
\begin{align}
    h_{a} &= EX_{seq}^{a}(ha) \\
    E_{a}^{I} &= EX_{emb}^{a}(E_a^I)
\end{align}
where $[\cdot]$ is the concatenation execution of two vectors or matrices. $EX_{seq}^{a}$ and $EX_{emb}^a$ are two extractor nets. The extractor net is a feed-forward network which can be defined as follows:
\begin{align}\label{eq:extract}
    h^t &= W^t \sigma(h^{t-1}) + b^t \quad (t=1,2,3...)
\end{align}
where $W^0,W^1,...,W^t$ and $b^0,b^1...,b^t$ are $2d \times 2d$ and $2d$ dimensional parameters need to learn ($d$ is the dimension of hidden states). $t$ is the number of layers in the extractor net. $\sigma$ is the activation function. In our work, we use $ReLU$ as the activation function. $h_0$ is the input of the feed-forward network (i.e., $[h_p, h_c]$ and $[E_p^I, E_c^I]$). Inspired by \cite{he16resnet}, we add the original input as the final representation:
\begin{equation}
    h^t = h^t + h^0
\end{equation}

Finally, the scores of all items can be computed as:
\begin{equation}
    \hat{y}_{a} = E_{a}^{I} h_{a}^\mathsf{T}
\end{equation}
which is similar to Eq. \ref{eq:dot}. We also use the weighted loss function $L_w$ to optimize the model. In conclusion, the loss function of the DACSR model can be written as:
\begin{equation}
    L = L_w(\hat{y}_a) + L_{Acc}(y, \hat{y}_p) + L_{Calib}(\hat{y}_c)
\end{equation}
where $\hat{y}_p$ and $\hat{y}_c$ are scores of all items generated by $f_p$ and $f_c$.

\section{Experiments}\label{sec:experiment}

\subsection{Dataset}
We adopted two commonly-used benchmark datasets to evaluate the performance of our model. The first one is \emph{Movelens-1m}\footnote{https://grouplens.org/datasets/movielens/1m}, which contains interaction logs of more than 6000 users and 3000 movies. The other is \emph{Tmall}\footnote{https://tianchi.aliyun.com/dataset/dataDetail?dataId=53}, which includes user behavior logs on an e-commerce platform. We retained the ``buy'' behaviors for the Tmall dataset. For both datasets, we sorted each user's behaviors according to the timestamp. We followed a 5-core and 20-core strategy for the Ml-1m and Tmall dataset that removes the users and items whose number of occurrences is less than 5 or 20. We also applied the leave-one-out evaluation protocol \cite{kang18sasrec}. The latest clicked item of a user belongs to the testing set, and the previous one of this item belongs to the validation set. The remaining sequences construct the training set. To augment the training data, we extended the user's sequence following \cite{li2017neural, wu2018session}. We set the maximum sequence length equal to 200 and 100 for the Ml-1m and Tmall dataset, respectively. The statistics are listed in Table \ref{table-statistic}.

\subsection{Comparison Models}
We selected the following methods as baselines:
\begin{itemize}
    \item \textbf{SASRec} \cite{kang18sasrec} is a self-attentive-based sequential recommendation model, and is a strong baseline. We apply the SASRec model as the sequence encoder for $f_p$ and $f_c$, and compare our model with SASRec.
    \item \textbf{CaliRec} \cite{steck18calibrated} is a post-processing model which re-ranks the results generated by the sequence encoder. It makes a trade-off between accuracy and calibration at each time step.
    \item \textbf{CaliRec-GC} \cite{silva21exploiting} utilizes an adaptive selection for the trade-off factor. The higher coverage of item attributes of the user historical sequence results in more consideration for calibration. 
\end{itemize}
\begin{table}[t]
\centering
\caption{Statistics of Datasets}
\begin{tabular}{cccc}
    \hline
     Statistics  & Ml-1m & Tmall\\
     \hline
     Number of users & 6,040 & 31,854\\
     Number of Items  & 3,883 & 58,343 \\
     Number of Training Sequences & 981,504 & 832,603  \\
     Number of Testing Sequences  & 6,040 & 31,854 \\
     Number of Attributes & 18 & 70 \\
     Average Length of Sequence & 164.50 & 28.13 \\
     \hline
\end{tabular}
\label{table-statistic}
\end{table}

\subsection{Evaluation Metrics}\label{sec:metrics}
We evaluate models from accuracy and calibration perspectives.  Following previous work\cite{wu2018session, li2017neural}, we use Recall and MRR as evaluation metrics to measure the recommendation accuracy of models. 
\begin{itemize}
    \item \textbf{Recall@K} (Rec@K) is a widely used metric in recommendation and information retrieval areas. Recall@K computes the proportion of correct items in the top-K items of the list.
    \begin{equation}
        Recall@K= \frac{1}{N} \sum_{s} 1(x_{n+1} \in RL_{s})
    \end{equation}
    where $1(\cdot)$ is an indication function whose value equals 1 when the condition in brackets is satisfied and 0 otherwise. $N$ is the number of testing cases.
    \item \textbf{MRR@K} is another important metric that considers the rank of correct items. The score is computed by the reciprocal rank when the rank is within K; otherwise the score is 0. 
    \begin{equation}
        MRR@K = \frac{1}{\|N\|} \sum_{s} \frac{1}{rank(x_{n+1}, RL_{s})}
    \end{equation}
\end{itemize}

To evaluate the effectiveness in terms of calibration, we adopt $C_{KL}$ which is a common metric used for calibrated recommendation \cite{steck18calibrated}. The $C_{KL}$ compares the consistency between two distribution:
\begin{equation}
    C_{KL}(RL, s) = \frac{1}{N}\sum_{s} \sum_{g \in G}p(g \mid s)\frac{p(g \mid s)}{\widetilde{q}(g \mid s)}
\end{equation}
The lower $C_{KL}$ value means we provide more calibrated recommendation lists. To avoid the division-by-zero error, we use $\widetilde{q}(g \mid s)=(1 - \alpha)(g \mid s)+\alpha p(g \mid s)$ to replace the original preference distribution $q(g \mid s)$. The value of $\alpha$ also equals to 0.01 according to \cite{steck18calibrated, Kaya19comparison, seymen21constrain, silva21exploiting}.

In addition, to better compare performances of our model and baseline models, we define the improvement as follows:
\begin{equation}
    Improv. = \frac{Metric_{DACSR} - Metric_{Baseline}}{Metric_{Baseline}}
\end{equation}
where $Metric$ can be any metric mentioned above.

\subsection{Experimental Setup}
We fixed the dimension of sequence representations and item embeddings equal to 64 for the DACSR model. For a fair comparison, we set the dimension of hidden states of the SASRec model to 128, so that the numbers of parameters are close. The number of layers in the extractor net $t$ of the DACSR model equals 2 for all datasets. We used the Adam \cite{adam} optimizer with the batch size of 256 and the learning rate of 0.001. We reported the performance under the model parameters with the optimal prediction accuracy on the validation set. We made hyper-parameter $\lambda = 0.5$ and $\tau = 1$ as the default setting, and analyzed their influence in following sections. To accelerate the training procedure, we initialized the parameters of sequence encoders used in our models by pre-trained parameters. For the top-K recommendation, we set $K=10$ and 20 which is a common setting.

\section{Results and Analysis}\label{sec:result}
In this section, we provided results and analysis of our work. In general, we aimed to answer the following research questions:
\begin{itemize}
    \item \textbf{RQ1} How the performances and efficiency of our DACSR model in achieving accurate and calibrated recommendation lists? 
    \item \textbf{RQ2} How the performances of our model change as the parameters change?
    \item \textbf{RQ3} How the modules of our DACSR model contribute to the performance improvement?
    \item \textbf{RQ4} How the distribution modification approaches work on the two datasets?
\end{itemize}

\begin{table}[tbp]
  \centering
  \caption{Performances of our models and baselines (best performances are marked in bold)}
    \begin{tabular}{c|c|c|c|c|c}
    \toprule
    Datasets & Metrics & SASRec & CaliRec & CaliRec-GC    & DACSR \\
    \midrule
    \multirow{6}[12]{*}{ML-1m} & Rec@10 & 0.2627 & 0.2636 & 0.2225 & \textbf{0.2811} \\
\cmidrule{2-6}          & MRR@10 & 0.1203 & 0.1101 & 0.0712 & \textbf{0.1267} \\
\cmidrule{2-6}          & $C_{KL}$@10 & 1.2385 & 0.9722 & \textbf{0.4553} & 1.0615 \\
\cmidrule{2-6}          & Rec@20 & 0.3613 & 0.3616 & 0.3258 & \textbf{0.3844} \\
\cmidrule{2-6}          & MRR@20 & 0.1271 & 0.1168 & 0.0784 & \textbf{0.1338} \\
\cmidrule{2-6}          & $C_{KL}$@20 & 0.8548 & 0.7322 & \textbf{0.3847} & 0.7262 \\
    \midrule
    \multirow{6}[12]{*}{Tmall} & Rec@10 & 0.1451 & 0.1464 & 0.1454 & \textbf{0.1517} \\
\cmidrule{2-6}          & MRR@10 & \textbf{0.0862} & 0.0846 & 0.0861 & 0.0857 \\
\cmidrule{2-6}          & $C_{KL}$@10 & 2.5004 & 2.0710 & 2.4139 & \textbf{2.0114} \\
\cmidrule{2-6}          & Rec@20 & 0.1749 & 0.1753 & 0.1751 & \textbf{0.1855} \\
\cmidrule{2-6}          & MRR@20 & \textbf{0.0883} & 0.0866 & 0.0882 & 0.0881 \\
\cmidrule{2-6}          & $C_{KL}$@20 & 2.1103 & 1.7943 & 2.0459 & \textbf{1.6240} \\
    \bottomrule
    \end{tabular}%
  \label{tab:overall}%
\end{table}%

\subsection{RQ1: Overall Performance}\label{sec:overall}
In this section, we answer the research question RQ1 about whether our model can provide calibrated and accurate recommendations. The performances of baselines and our model are listed in Table \ref{tab:overall}, where the best performance is marked in bold.

\textbf{Recommendation Accuracy} We first analyze the performance from the perspective of accurate recommendation (i.e., Rec@K and MRR@K). In general, on both datasets, our model achieves the best prediction accuracy in terms of Recall and MRR. By considering calibration, users' preference distributions are incorporated. In addition, our model decoupled the two objectives by two sequence encoders and aggregated their outputs. Therefore, the preference distribution contributed to the prediction of the next item, leading to the improvement of accuracy. For example, on the Ml-1m dataset, the Recall and MRR of our model are higher than the original SASRec model (e.g., 0.1338 v.s. 0.1271 in terms of MRR@20). In contrast, the post-processing-based models fragmented the relationship between accuracy and calibration and therefore resulted in a reduction in accuracy. For example, on the Ml-1m dataset, the MRR@20 of our model is 0.1338, while it is 0.1168 and 0.0784 for CaliRec and CaliRec-GC model, with the improvement of $18.67\%$ and $74.66\%$, respectively. On the Tmall dataset, the CaliRec model also decreases the prediction accuracy. 

\textbf{Calibrated Recommendation} Our model can provide more calibrated recommendation lists compared to the original sequential recommendation model. On both datasets, $C_{KL}@10$ and $C_{KL}@20$ of our model are lower than the original SASRec model. For example, the $C_{KL}@20$ of our model is 0.7262, which is $15.04\%$ better than the 0.8548 of the SASRec model. On the Tmall dataset, our model also achieves a $23.04\%$ improvement in terms of $C_{KL}@20$. Compared to the post-processing-based CaliRec model, our model still achieves competitive performances in terms of calibration. For example, on the ML-1m dataset, the performances of $C_{KL}@20$ of our model and the CaliRec model are 0.7262 and 0.7322. On the Tmall dataset, our model achieves an improvement of $9.49\%$ in terms of $C_{KL}@20$. The comparisons show the ability of our model to achieve better accuracy while obtaining competitive performances of calibration compared to the post-processing-based models. As for the possible reasons, on the one hand, the proposed loss function calibrated the preference distribution of items with the highest scores to the historical preference distribution. On the other hand, the decoupled-aggregated framework ensures accuracy when improving the calibration.

We also observe that the CaliRec-GC model performs differently on the two datasets. On the Ml-1m dataset, the CaliRec-GC model achieves the lowest $C_{KL}$ value among all models, including our proposed model (e.g., 0.3847 v.s. 0.7262 of $C_{KL}@20$). While on the Tmall dataset, the CaliRec-GC model cannot provide calibrated recommendation lists. For example, the $C_{KL}@20$ of CaliRec-GC is $2.0459$, which is close to the original SASRec model. We think this phenomenon results from two aspects. On the one hand, the number of item attributes of the Tmall dataset is much more than that of the Ml-1m dataset. The Ml-1m dataset contains 18 different item attributes, while the Tmall dataset has 70 attributes. On the other hand, the average length of the user behavior sequence of the Tmall dataset is less than the Ml-1m dataset, as shown in Table \ref{table-statistic}. The shorter sequence and larger item attributes set lead to the lower coverage of item attributes. The CaliRec-GC model adopts an adaptive selection of the trade-off factor $\lambda$ for calibration based on the coverage of item attributes. The greater coverage leads to the higher value of $\lambda$. Therefore, it performs best in terms of calibration on the Ml-1m dataset, and almost does not work on the Tmall dataset.

The differences between the two datasets also lead to the different calibration performances of the two datasets. In general, performances of $C_{KL}$ on the Ml-1m dataset are better than the Tmall dataset. For example, the $C_{KL}@20$ of our DACSR model is 1.6240 on the Tmall dataset, which is much higher than the 0.7262 on the Ml-1m dataset. This is similar for the original SASRec model, with 2.1103 v.s. 0.8548 on the two datasets. A possible reason is that the lower coverage of item attributes mentioned above results in the higher divergence. The large amount of 0 in $p_(s)$ makes it difficult in achieving calibration, especially with the concern of accuracy.

\textbf{Time Consumption} In addition, we also compared the response time of our model against the post-processing-based CaliRec model and the original SASRec model. We focused on the average time required to generate the recommended list for each sequence. For the SASRec model, we reported the time consumption when the dimension of hidden states equals to 64 and 128 (namely $SASRec_{D64}$ and $SASRec_{D128}$). This is because the size 64 is the setting of each sequence encoder of our DACSR model. We conducted experiments on the same device, and removed the GPU acceleration for a fair comparison. The performance is listed in Table \ref{tab:time}.

Compared to the original SASRec model, our model needs more computation. For example, on the Ml-1m dataset, the time consumption for each sequence of the $SASRec_{D64}$ model is 2.01 $\times 10^{-4}$ seconds, which is approximately half of our DACSR model. This is because it incorporates two SASRec encoders and an extraction net, which is more complex than the single SASRec model. In contrast, our model can provide more accurate and calibrated recommendations than the original SASRec model. The $SASRec_{D128}$ requires more time than $SASRec_{D64}$ because it contains a larger scale of parameters. Compared to the CaliRec model, our model costs much less time to generate recommendation lists. For a single sequence, our model only needs 4.24 and 5.76 $\times 10^{-4}$  seconds on the Ml-1m and Tmall datasets, respectively. In contrast, the CaliRec requires approximately 0.06 seconds for a sequence, which needs 200 times more time than our model. This is because the CaliRec model needs an extra ranking stage. The original SASRec model provided scores of all items, and selected the top-100 items with the highest scores. The post-processing-based CaliRec model then re-ranks the top-100 items with $K$ steps ($K$ stands for the top-K recommendation). At each step, it computes the gains of the candidate items when they are added to the recommendation list. However, our model follows an end-to-end framework only with a sorting stage to select top-K items after the scores of all items are computed. Therefore, our model obtains better performance and requires less time consumption than the CaliRec model for each sequence. 

\begin{table}[tbp]
  \centering
  \caption{Average time consumption for each sequence.}
    \begin{tabular}{c|cc}
    \toprule
    Response time ($10^{-4}$s) & \multicolumn{1}{l}{Ml-1m} & \multicolumn{1}{l}{Tmall} \\
    \midrule
    $SASRec_{D64}$ & 2.01 & 1.32 \\
    $SASRec_{D128}$ & 3.28 & 1.96 \\
    CaliRec & 580.15 & 668.69 \\
    DACSR & 4.24 & 5.76 \\
    \bottomrule
    \end{tabular}%
  \label{tab:time}%
\end{table}%

\textbf{Generalization of DACSR Model} We are also interested in whether our model is also effective when the sequence encoder changes. We incorporated the GRU4Rec model \cite{hidasi2015session} as the sequence encoder, which is also a widely used sequential recommendation model. The experimental settings were same to the previous section with $\lambda=0.5$. We use DACSR(G) to denote our DACSR model which takes GRU4Rec as the sequence encoder, and CaliRec(G) to denote the post-processing-based CaliRec model with candidates provided by the GRU4Rec model. The performances are listed in Table. \ref{tab:gru4rec}.

As shown in the table, our model can still achieve calibrated and accurate recommendation lists when we use GRU4Rec as the sequence encoder. On the Ml-1m dataset, the performances of $C_{KL}@20$ are 0.6840 and 0.8356 of our DACSR(G) model and the GRU4Rec model, respectively. On the Tmall dataset, our model also obtains an improvement of $21.95\%$ in terms of calibration. Toward recommendation accuracy, the performances of our model are still better than the original GRU4Rec model. The improvement is not as great as the DACSR model with the SASRec sequence encoder. We think that this is because the ability to model sequences of GRU4Rec is worse than that of the SASRec model. The SASRec model with self-attention mechanisms can better find the user's preference and represent the sequence. The CaliRec(G) model also sacrificed the ranking performance to improve the calibration, which is similar to the CaliRec model with the SASRec model. Compared to the CaliRec(G) model, our DACSR(G) model can also achieve better performance in terms of accuracy and calibration, as listed in Table. \ref{tab:gru4rec}. The performance comparisons indicate that our model can be used for other basic sequence encoders, which is not specifically designed for the SASRec model.

\begin{table}[htbp]
  \centering
  \caption{Performances of our DACSR(G) model with the GRU4Rec sequence encoder (best performances are marked in bold).}
    \begin{tabular}{c|c|c|c|c}
    \toprule
    Dataset & Method & Rec@20 & MRR@20 & $C_{KL}$@20 \\
    \midrule
    \multirow{3}[6]{*}{Ml-1m} & GRU4Rec & 0.346 & 0.1142 & 0.8356 \\
\cmidrule{2-5}          & CaliRec(G) & 0.3454 & 0.1105 & 0.7013 \\
\cmidrule{2-5}          & DACSR(G) & \textbf{0.3472} & \textbf{0.1161} & \textbf{0.6840} \\
    \midrule
    \multirow{3}[6]{*}{Tmall} & GRU4Rec & 0.1724 & 0.0863 & 2.2123 \\
\cmidrule{2-5}          & CaliRec(G) & 0.1732 & 0.0846 & 1.8597 \\
\cmidrule{2-5}          & DACSR(G) & \textbf{0.1750} & \textbf{0.0878} & \textbf{1.7266} \\
    \bottomrule
    \end{tabular}%
  \label{tab:gru4rec}%
\end{table}%

\subsection{RQ2: Parameter Influence}

In this section, we answer the research question RQ2 about the influence of hyperparameters. Specifically, we investigate the two hyperparameters $\lambda$ and $\tau$ (see Sec. \ref{sec:loss}). The trade-off parameter $\lambda$ controlled the importance of calibration during the optimization stage of our model. The parameter $\tau$ reshaped the predicted scores of all items, which can affect the computation of the calibration loss function.

\begin{figure}[tbp]
\centering
\subfigure[Rec@20 on the Ml-1m dataset]{
\includegraphics[width=5cm]{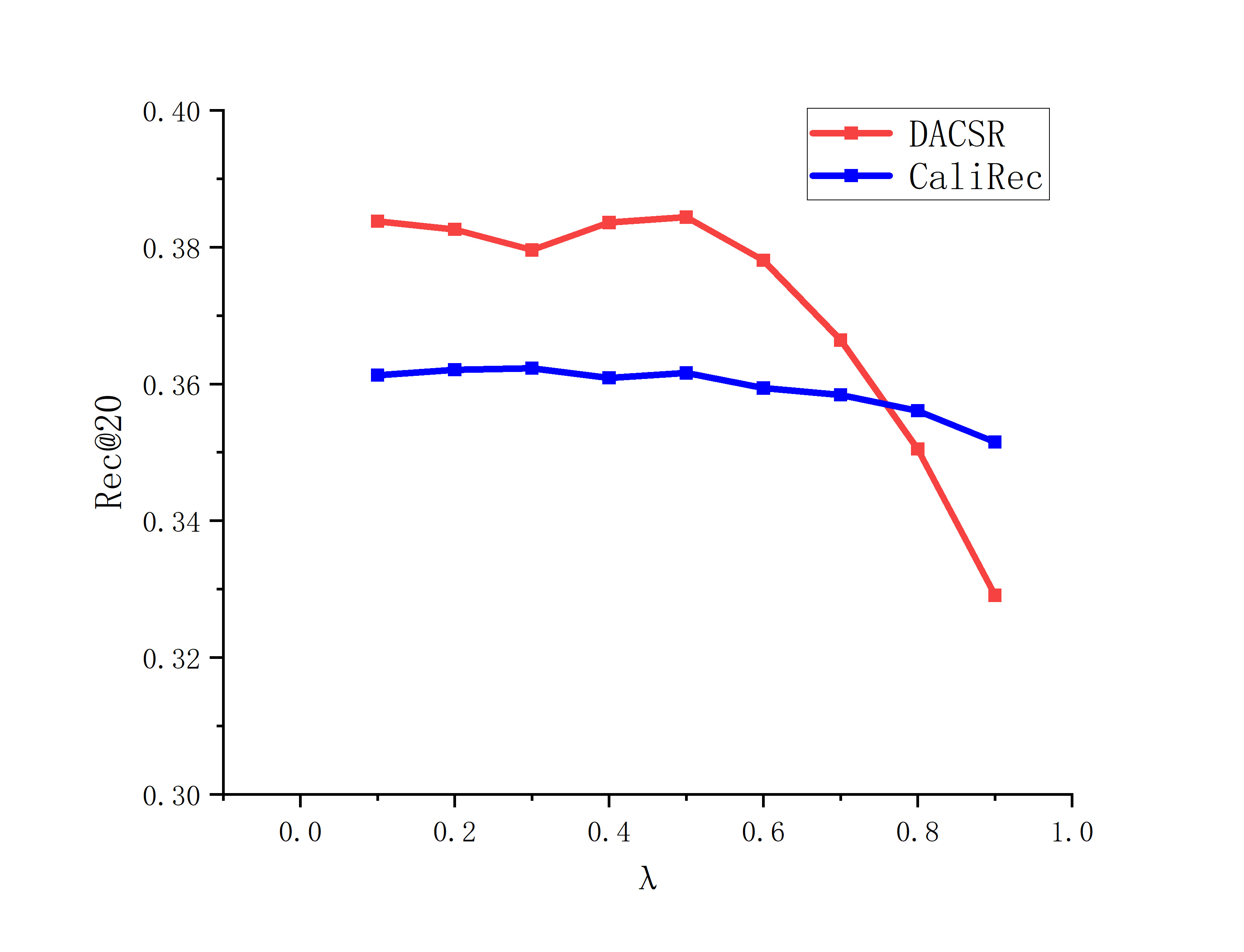}
\label{subfig:lamda-rec-ml}
}
\quad
\subfigure[Rec@20 on the Tmall dataset]{
\includegraphics[width=5cm]{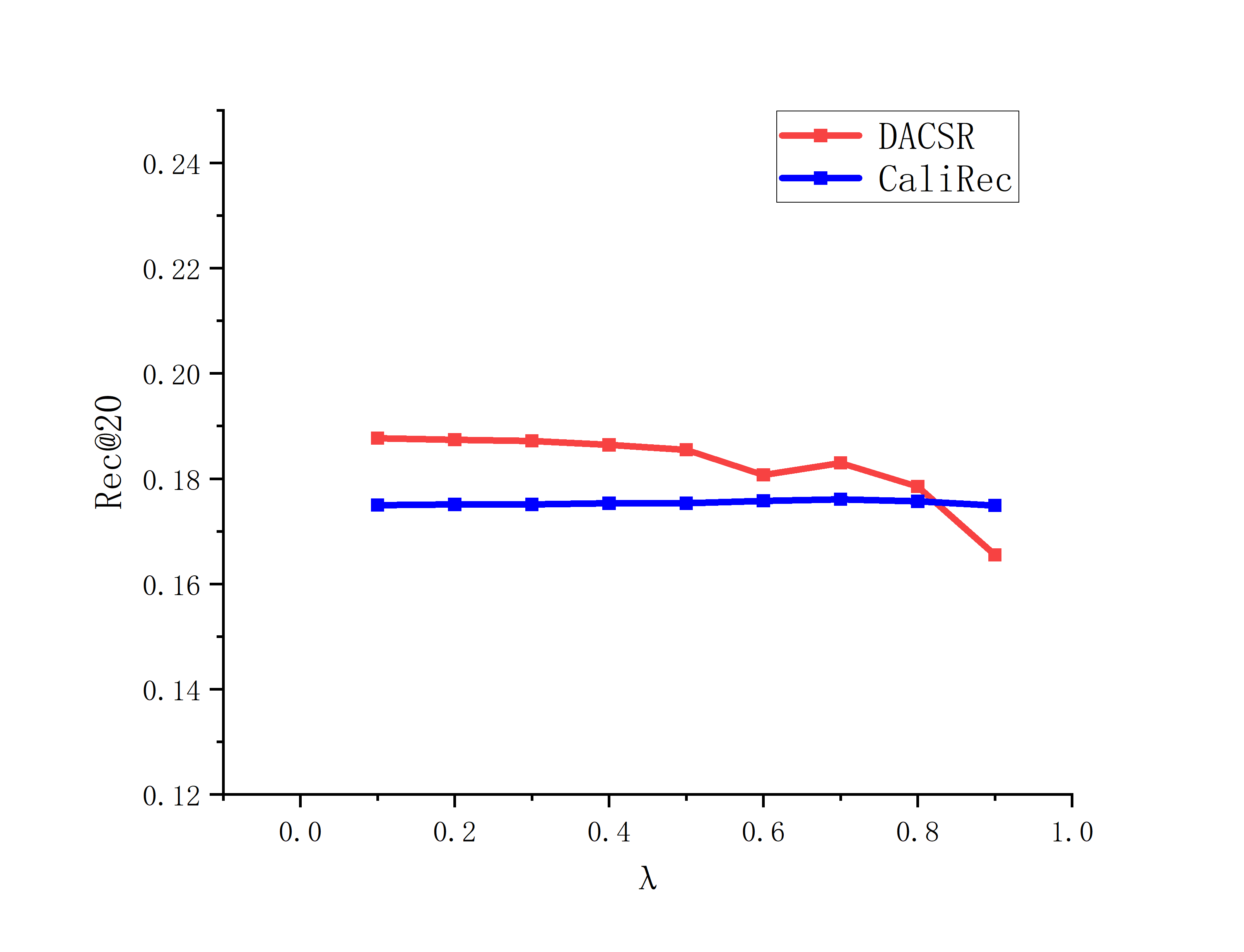}
\label{subfig:lamda-rec-tmall}
}
\subfigure[MRR@20 on the Ml-1m dataset]{
\includegraphics[width=5cm]{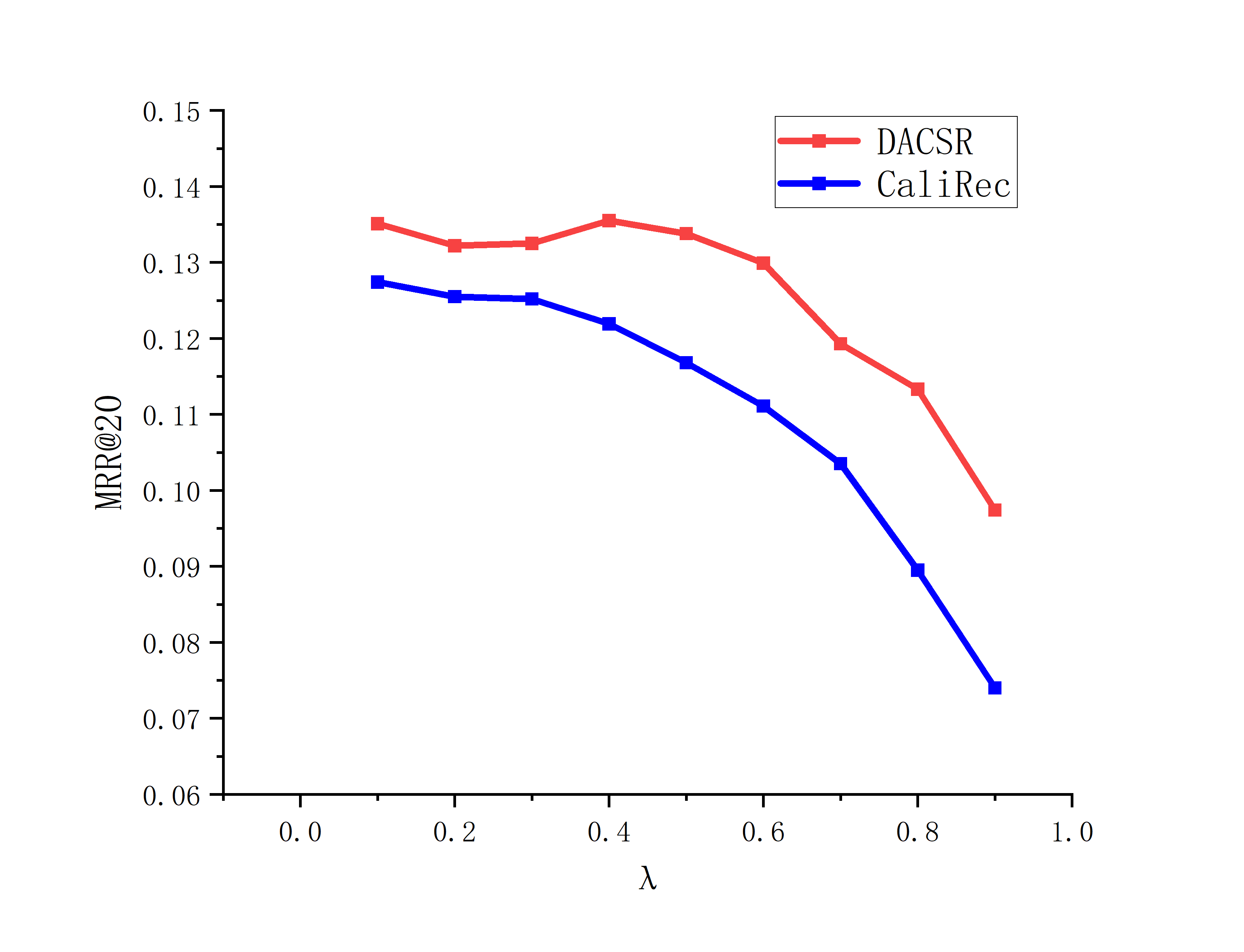}
\label{subfig:lamda-mrr-ml}
}
\quad
\subfigure[MRR@20 on the Tmall dataset]{
\includegraphics[width=5cm]{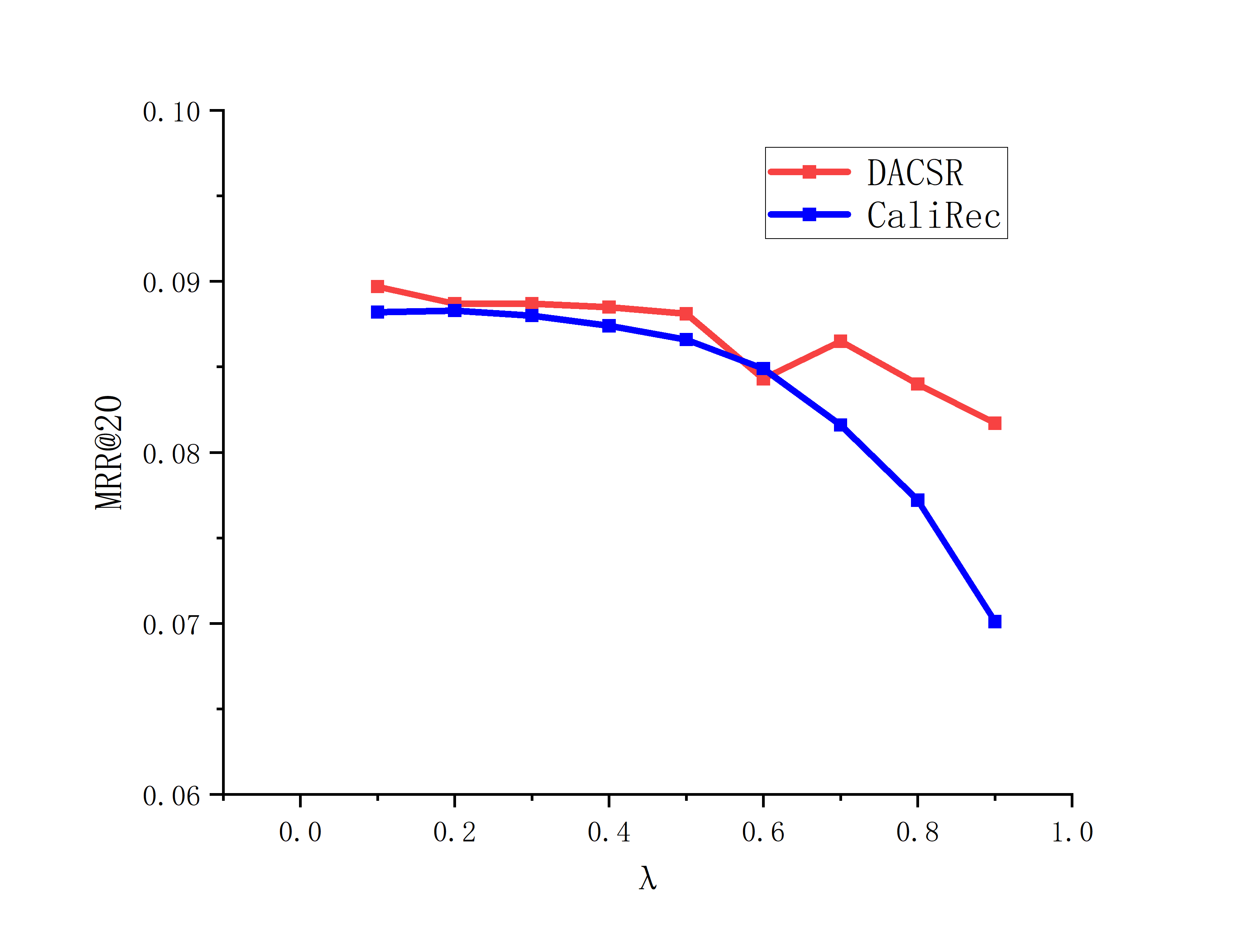}
\label{subfig:lamda-mrr-tmall}
}
\subfigure[$C_{KL}@20$ on the Ml-1m dataset]{
\includegraphics[width=5cm]{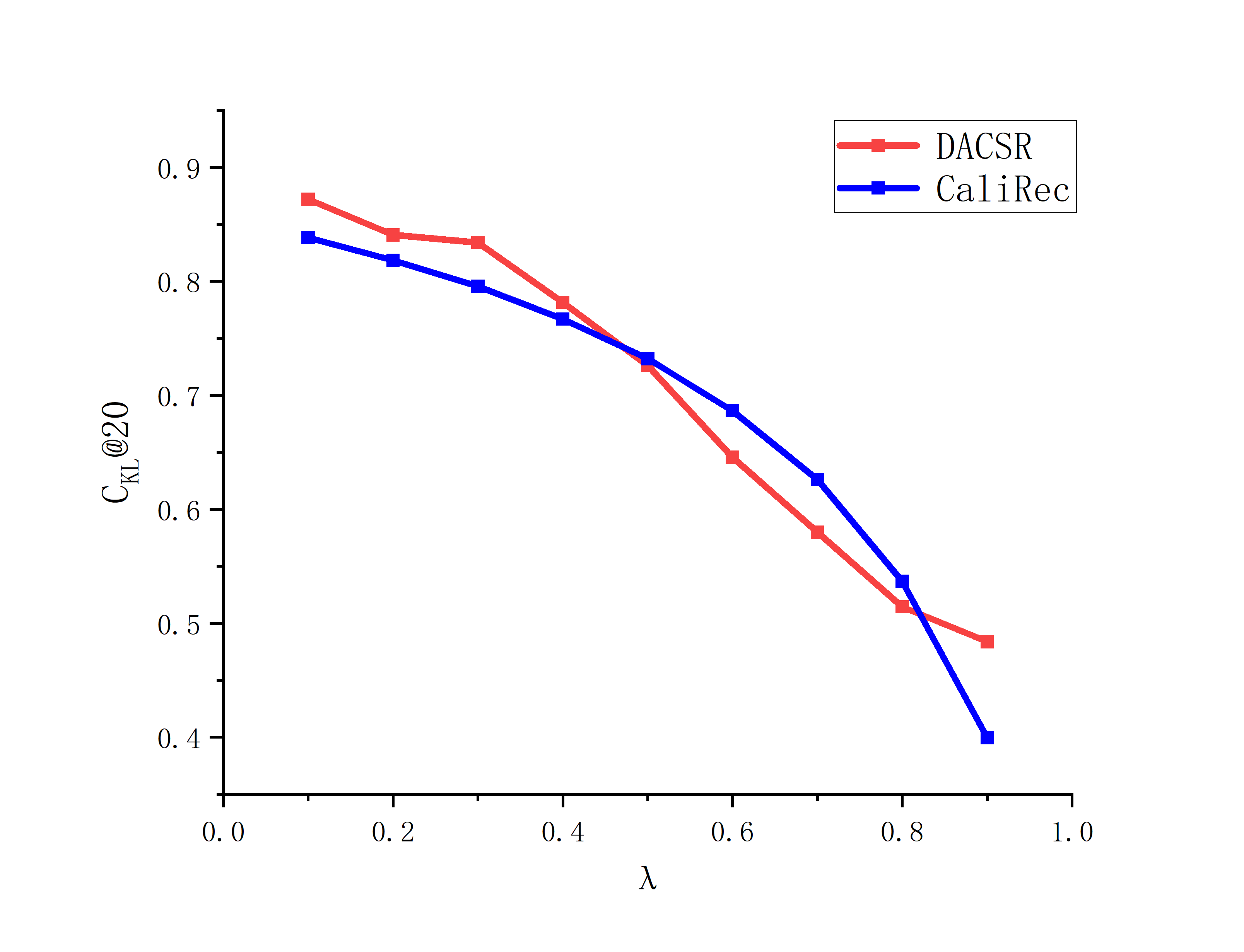}
\label{subfig:lamda-ckl-ml}
}
\quad
\subfigure[$C_{KL}@20$ on the Tmall dataset]{
\includegraphics[width=5cm]{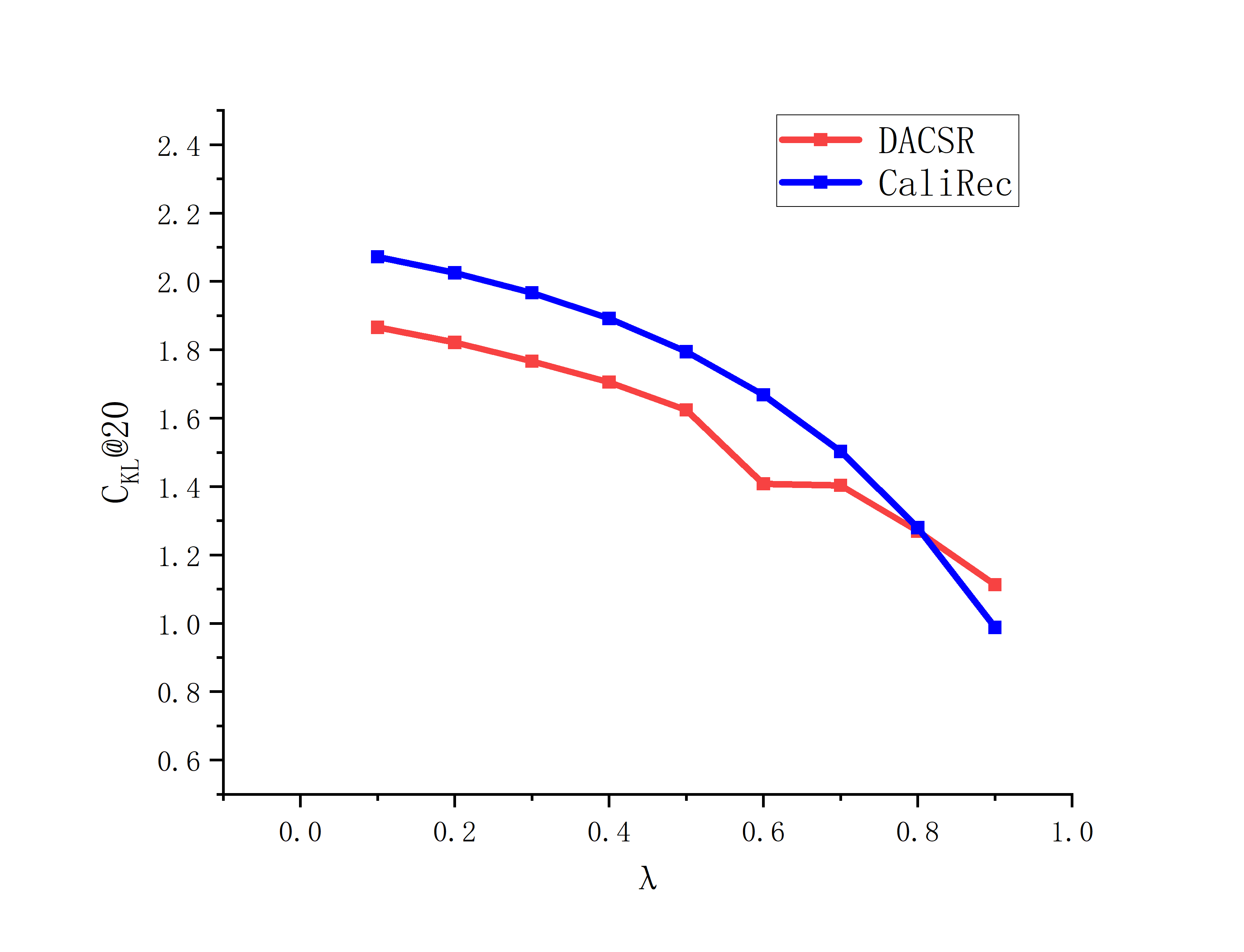}
\label{subfig:lamda-ckl-tmall}
}

\caption{Performance comparison when $\lambda$ changes.}
\label{fig:lamda}
\end{figure}

\subsubsection{Trade-off Factor $\lambda$}
We first analyze the influence of hyperparameter $\lambda$ by changing it from 0.1 to 1.0. Since the CaliRec model also required a parameter to control the importance of calibration and accuracy which was similar to our model, we displayed the changes in the performance of our DACSR model and compared it with the CaliRec model simultaneously. The performances on the two datasets are shown in Fig. \ref{fig:lamda}, where red lines and blue lines represent our DACSR model and the CaliRec model, respectively. 

In general, a greater value of $\lambda$ leads to a lower recommendation accuracy and higher calibration. When $\lambda$ has a high value (e.g., $\lambda=0.7$), the performance of Rec@20 of our model decreases greatly, while the CaliRec model does not decrease as large as our model. A possible reason is the difference in the process of generating recommendation lists. The CaliRec re-ranks the candidate list generated by the SASRec model whose size is 100 and finally selects top-K (K = 10 or 20) items as the final recommendation list. However, our model computes scores of all items, and directly selects top-K items. For some sequences, the users' next items achieve relatively lower ranks. When considering more for calibration, these items may be replaced by items whose attributes are consistent with the historical distributions but not sequential correlated to the sequences. While for post-processing methods, the range of items for re-ranking is narrowed. Therefore, it can preserve the performance of Rec@20. In contrast, our model outperforms CaliRec when considering ranking performances in terms of MRR. As shown in Fig. \ref{subfig:lamda-mrr-ml} and \ref{subfig:lamda-mrr-tmall}, the MRR@20 of our model is always higher than the CaliRec model. This is because although target items may be replaced, these items do not contribute to the ranking performance greatly. In contrast, our model improves the ranking of the target items for most of the sequences, resulting in an improvement in overall ranking performance.

In terms of calibration, our model achieves better performances compared to the CaliRec model under the same $\lambda$ in general. On the Tmall dataset, the performances of calibration of our DACSR model are better than the CaliRec model during the major changing process of $\lambda$. On the Ml-1m dataset, the performances are close. Our model performs better than the CaliRec model when $\lambda$ is greater than $0.5$. When $\lambda$ is close to 1, our DACSR model cannot perform as well as the CaliRec model. A possible reason is that the CaliRec model utilized a greedy-based ranking strategy so that it can select the most calibrated recommendations from top-100 items when $\lambda$ is close to 1. In contrast, the deep learning-based optimization strategy always has a gap in fitting the target distribution and therefore does not perform as well as the CaliRec model at very high values of $\lambda$. However, a large $\lambda$ leads to lower recommendation accuracy. Though it achieves better calibration, it is not suggested to use high value of $\lambda$ because accurately predicting the user's next behavior is still an important concern.

\begin{figure}[tbp]
\centering
\subfigure[Performance change on the Ml-1m dataset]{
\includegraphics[width=5cm]{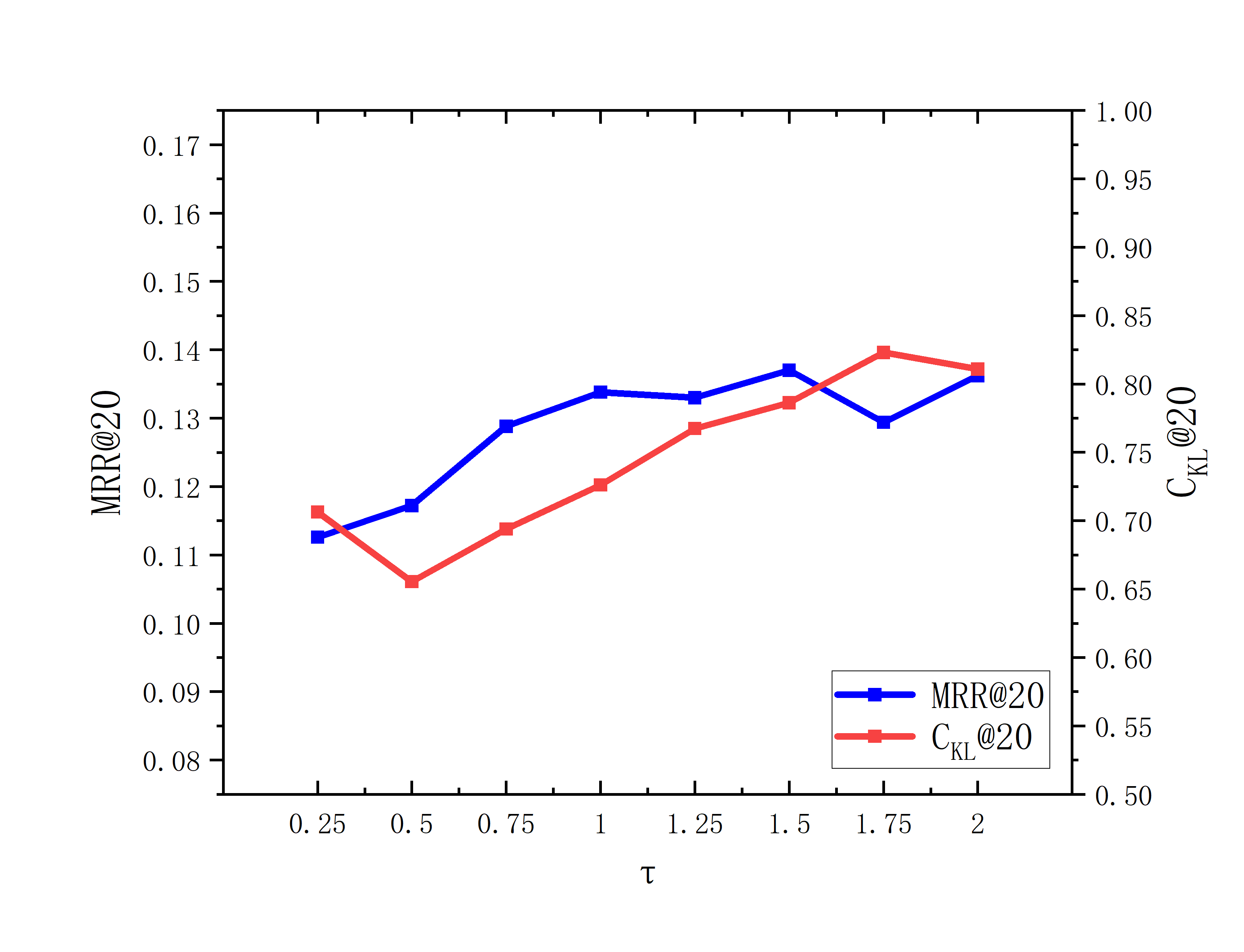}
\label{subfig:tau-ml}
}
\quad
\subfigure[Performance change on the Tmall dataset]{
\includegraphics[width=5cm]{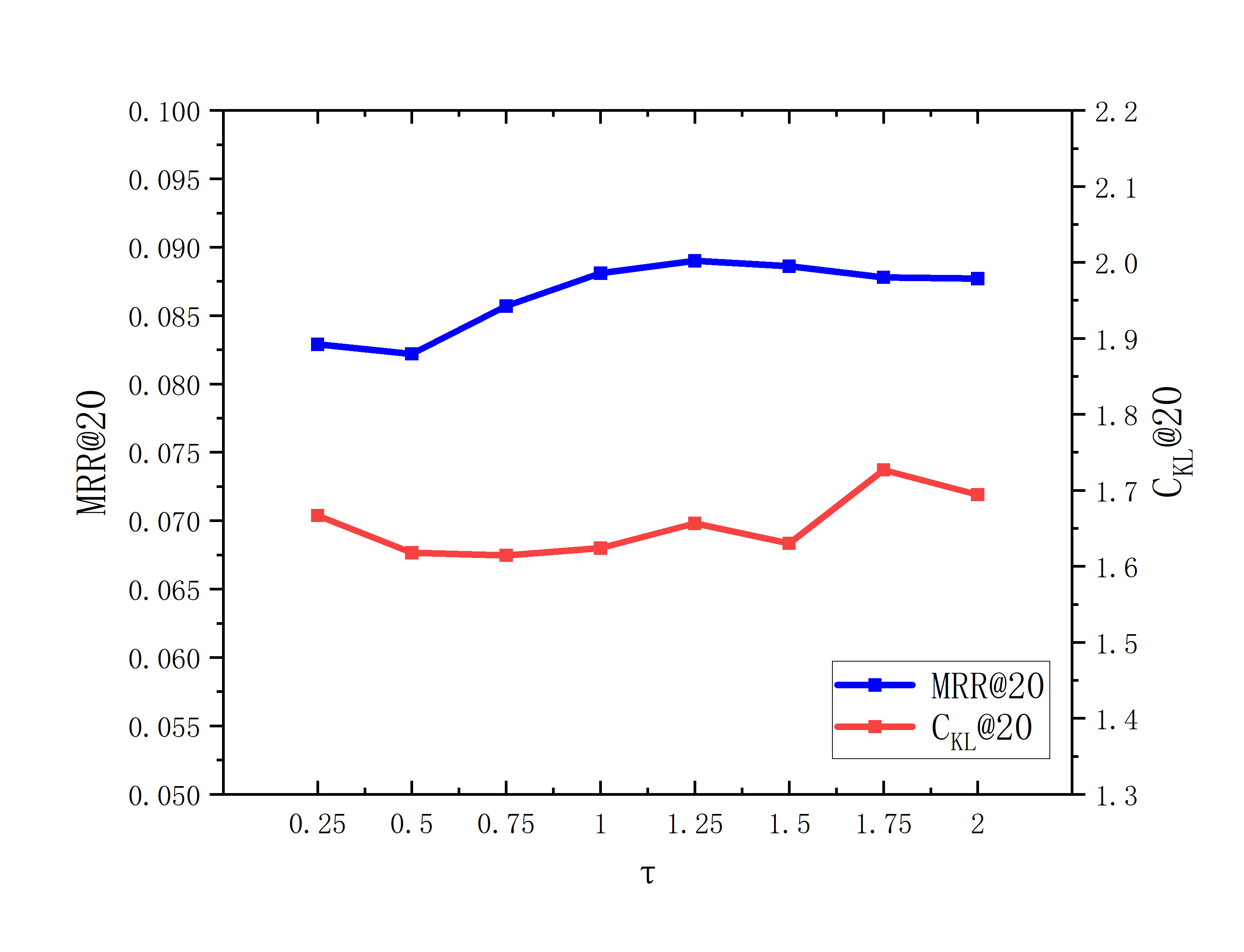}
\label{subfig:tau-tmall}
}
\caption{Performance comparison when $\tau$ changes.}
\label{fig:tau}
\end{figure}

\subsubsection{Temperature Factor $\tau$}

In this section, we analyze the influence of parameter $\tau$, which controls the sharpness of the distribution in the calibration loss function. We tune $\tau$ in $\{0.25, 0.5, 0.75, 1.0, 1.25, 1.5, 1.75, 2.0\}$, and show the performances in Fig. \ref{fig:tau}. Red lines represent the performance of $MRR@20$ and correspond to the right axis, and green lines stand for the $C_{KL}@20$ performance which follows the left axis. 

On the two datasets, a relatively lower value of $\tau$ can achieve a better performance of calibration and lower accuracy in general. For example, on the Ml-1m dataset, the $C_{KL}@20$ performances are 0.6466 and 0.6915 when $\tau$ equals to 0.25 and 1.0 respectively. This is because a lower value of $\tau$ amplifies the scores of top items, and other items are ignored because their scores are normalized to 0. The lower value of $\tau$ increases the calibration, but decreases the recommendation accuracy, as shown by the blue lines in Fig. \ref{fig:tau}. In contrast, a higher value of $\tau$ ($\tau > 1$) causes a negative impact on calibration, because scores of all items are normalized to close values (i.e., $1 / {\mid I \mid}$ for all items). Therefore, no useful information for calibration can be propagated to the model. 

\subsection{RQ3: Ablation Studies}

To answer the research question RQ3, we conduct ablation experiments in this section by comparing our model with two variants. The first variant is the original SASRec model optimized by the loss function $L_w$. We aim to investigate the performance of a single sequence encoder optimized by both accuracy and calibration. We also reported the performances when the dimension of hidden states equals to 64 and 128 (namely $SASRec^{L_w}_{D64}$ and $SASRec^{L_w}_{D128}$). In addition, we directly add the extractor nets to the $SASRec^{L_w}_{D128}$ model, namely ``$SASRec^{L_w}_{D128}EX$''. The other one is the direct concatenation of sequence representations and item embedding matrices without extraction nets (namely DACSR-C). We compare our model with these variants by setting $\lambda=0.5$. The performances are listed in Table \ref{tab:ablation}. In general, our DACSR model obtains the best performance in terms of recommendation accuracy and calibration.

The effectiveness of our designed loss function for calibration can be reflected by the performance of $SASRec^{L_w}_{D64}$ and $SASRec^{L_w}_{D128}$. By applying the loss function $L_w$, the SASRec model is able to provide more calibrated recommendation lists than the original SASRec model only optimized by $L_{Acc}$. For example, the $C_{KL}@20$ of SASRec on the Tmall dataset decreases from 2.1103 to 1.8845. Also, it achieves close performance compared to our DACSR model on the Ml-1m dataset in terms of calibration. However, a larger space of parameters does not mean better performances in terms of both accuracy and calibration. Although the $SASRec^{L_w}_{D128}$ model improved the accuracy and achieved close performance of calibration on the Ml-1m dataset, it obtained less calibrated recommendations on the Tmall dataset. The increased parameter size of the model leads to better learning ability, but it still does not optimize both objectives well simultaneously. 
In addition, the performances of the $SASRec^{L_w}_{D128}EX$ model are also worse than the DACSR model.
We believe that this is because an individual model that improves performance in one area may negatively affect performance in the other because their parameters are shared.

\begin{table}[tbp]
  \centering
  \caption{Performance comparisons between our DACSR model and its variants}
    \begin{tabular}{c|c|ccccc}
    \toprule
    Dataset & Metrics & $SASRec^{L_w}_{D64}$ & $SASRec^{L_w}_{D128}$ & $SASRec^{L_w}_{D128}EX$ & DACSR-C & DACSR \\
    \midrule
    \multirow{6}[2]{*}{Ml-1m} & Rec@10 & 0.2551  & 0.2730  & 0.2715  & 0.2710  & \textbf{0.2811 } \\
          & MRR@10 & 0.1088  & 0.1206  & 0.1213  & 0.1187  & \textbf{0.1267 } \\
          & CKL@10 & 1.0780  & 1.0765  & 1.0835  & 1.0909  & \textbf{1.0615 } \\
          & Rec@20 & 0.3599  & 0.3657  & 0.3748  & 0.3719  & \textbf{0.3844 } \\
          & MRR@20 & 0.1161  & 0.1269  & 0.1283  & 0.1257  & \textbf{0.1338 } \\
          & CKL@20 & 0.7281  & \textbf{0.7253 } & 0.7315  & 0.7433  & 0.7262  \\
    \midrule
    \multirow{6}[2]{*}{Tmall} & Rec@10 & 0.1508  & 0.1482  & 0.1462  & 0.1464  & \textbf{0.1517 } \\
          & MRR@10 & 0.0854  & 0.0862  & \textbf{0.0864 } & 0.0813  & 0.0857  \\
          & CKL@10 & 2.1490  & 2.2499  & 2.3119  & \textbf{1.9816 } & 2.0114  \\
          & Rec@20 & 0.1830  & 0.1787  & 0.1777  & 0.1804  & \textbf{0.1855 } \\
          & MRR@20 & 0.0876  & 0.0884  & \textbf{0.0886 } & 0.0837  & 0.0881  \\
          & CKL@20 & 1.7810  & 1.8845  & 1.9322  & \textbf{1.5960 } & 1.6240  \\
    \bottomrule
    \end{tabular}%
  \label{tab:ablation}%
\end{table}%

Our DACSR model outperforms the variants, demonstrating the effectiveness of the decoupled-aggregated framework. For example, on the Ml-1m dataset, the MRR@20 of our DACSR model is 0.1338, while it is 0.1269 for the $SASRec^{L_w}_{D128}$ model, and the performances of calibration are close (0.7262 v.s. 0.7253). On the Tmall dataset, our DACSR model can achieve competitive recommendation accuracy, and provide more calibrated recommendations (e.g., 1.6240 v.s. 1.8845 in terms of $C_{KL}@20$). Compared to the $SASRec^{L_w}_{D128}$ model which shares parameters for two objectives, the decoupled-aggregated framework can achieve better performance. We believe that such a framework can learn the information of two objectives and combine them to obtain better representations of sequences and items. The DACSR-C model removed the extraction net and directly concatenated the representations of sequences and items from two sequence encoders. It obtained worse performance than the DACSR model, showing the importance of the extraction net. On the Ml-1m dataset, the $C_{KL}@20$ of the DACSR model is 0.7262, which is slightly better than the 0.7433 of the DACSR-C model. But the recommendation accuracy of the DACSR model is higher than the DACSR-C model (e.g., 0.1338 v.s. 0.1257 in terms of MRR@20). On the Tmall dataset, our DACSR model also obtains better performances in terms of accuracy, and close performance of calibration. The extraction net takes the concatenation of sequence/item representations as inputs, and provides more suitable representations for the two objectives.


\subsection{RQ4: Distribution Modification}
In this section, we answer the research question RQ4 about the effectiveness of the proposed distribution modification approaches. We proposed the modified distribution $p_d(s)$ and $p_m(s)$ to further improve the diversity and mitigate the imbalanced interest problem. These approaches are related to the diversity. Therefore, we adopted the ILD metric with Jaccard similarity to measure the diversity of the recommendation list:
\begin{equation}
    ILD(RL_s) = \frac{2}{\mid RL_s \mid (\mid RL_s \mid - 1)} \sum_{(i,j \in RL_s)} (1 - \frac{\mid Attr_i \cap Attr_j \mid}{\mid Attr_i \cup Attr_j \mid})
\end{equation}
where $Attr_i$ is the item attribute set that the item $i$ has, and $RL_s$ is the generated recommendation list for sequence $s$. The larger ILD value represents the higher diversity of the recommendation list. We set the factor $\tau_{div}=0.5$ and 2 for the distribution $p_d(s)$ and $p_m(s)$, respectively. 
\begin{table}[tbp]
  \centering
  \caption{Performances of calibration and diversity (best performances are marked in bold)}
    \begin{tabular}{c|c|c|c|c|c}
    \toprule
    Datasets & Models & ILD@10 & $C_{KL}$@10 & ILD@20 & $C_{KL}$@20 \\
    \midrule
    \multirow{3}[6]{*}{Ml-1m} & SASRec & 0.6499 & 1.2385 & 0.6677 & 0.8548 \\
\cmidrule{2-6}          & DACSR-$p(s)$ & 0.6654 & \textbf{1.0615} & 0.6789 & \textbf{0.7262} \\
\cmidrule{2-6}          & DACSR-$p_d(s)$   & \textbf{0.7012} & 1.1347 & \textbf{0.7123} & 0.7649 \\
    \midrule
    \multirow{3}[6]{*}{Tmall} & SASRec & 0.7086 & 2.4871 & 0.7405 & 2.1092 \\
\cmidrule{2-6}          & DACSR-$p(s)$ & 0.6714 & \textbf{2.0114} & 0.7045 & \textbf{1.624} \\
\cmidrule{2-6}          & DACSR-$p_d(s)$   & \textbf{0.7419} & 2.2662 & \textbf{0.7725} & 1.8412 \\
    \bottomrule
    \end{tabular}%
  \label{tab:div}%
\end{table}%
We first listed the performances of our DACSR model with the raw historical preference distribution (namely DACSR-$p(s)$) and the modified preference distribution for diversity (namely DACSR-$p_d(s)$) along with the original SASRec model in Table \ref{tab:div}.

On the two datasets, the diversity of our model is improved by sacrificing the calibration. For example, on the Ml-1m dataset, the performances of $ILD@10$ are 0.7012 and 0.6654 for the normalized distribution $p_d(s)$ and the original distribution $p(s)$, respectively. However, the $C_{KL}@10$ increases from 1.0615 to 1.1347, which means the ability of calibration of our model is weakened. On the Tmall dataset, the performance comparisons are similar. This is because applying the normalized distribution amplifies the effect of item attributes that the user did not interacted in the behavior sequence. Though it does not largely affect the true distribution $p(s)$, it deviates from the calibration to a certain degree.

We observe that our DACSR model performed differently on the two datasets. On the Ml-1m dataset, the diversity is higher for our DACSR model than the original SASRec model, while it is totally different on the Tmall dataset. On the Tmall dataset, our DACSR model achieves worse performance in terms of diversity (e.g., 0.6714 v.s. 0.7086 of DACSR and SASRec model). It is possibly due to the difference between two datasets. On the Ml-1m dataset, the coverage of item attributes is higher than the Tmall dataset. Users have more historical behaviors than the Tmall dataset. On the Tmall dataset, users always interacted with several types of items, so that the $p(g \mid s)$ score of most attributes equals to 0. The limited interest areas resulted in less diversified recommendation lists under the calibration objective. 

\begin{figure}[tbp]
\centering
\includegraphics[width=12cm]{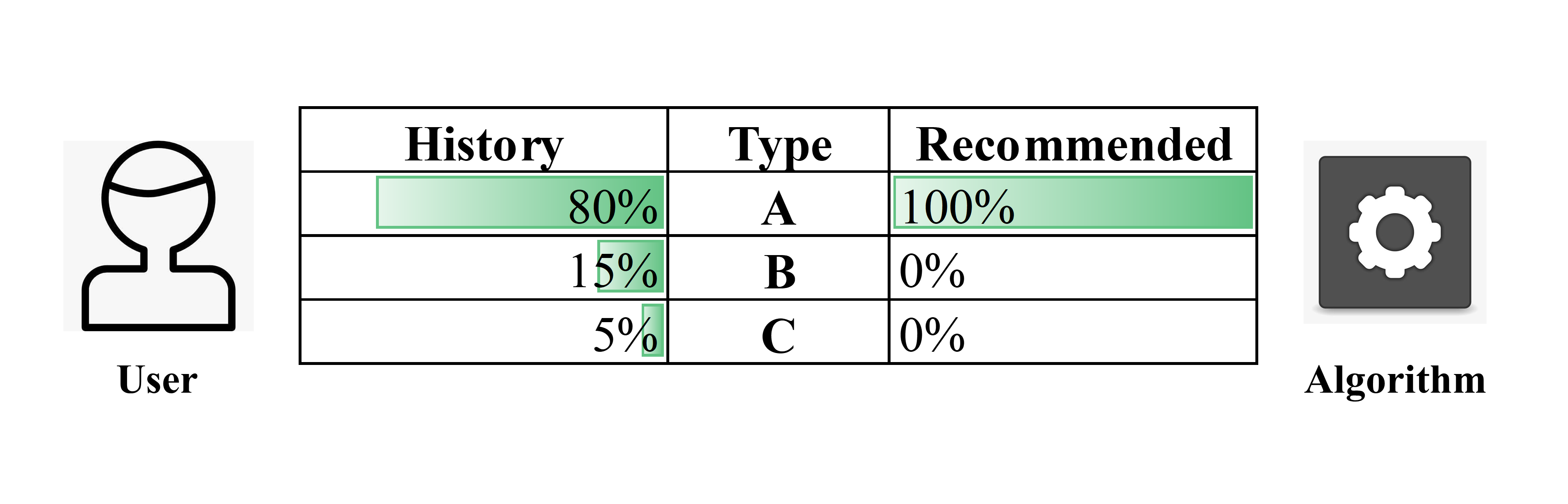}
\caption{Illustration of amplified main interest.}
\label{fig:imbalance}
\end{figure}

We also investigate the imbalanced interest problem. We find that imbalanced interests are amplified on the Tmall dataset. As illustrated in Fig. \ref{fig:imbalance}, the attribute $A$ occupies the 80\% of the sequence, while attribute $B$ and $C$ only account for the 20\%. For such an imbalanced distribution, our DACSR model amplifies the major interest, as shown in Fig. \ref{fig:imbalance}. The recommended list only contains items with attribute $A$. This gives our model a negative impact in terms of diversity. By applying the distribution $p_m(s)$, the diversity increases and the calibration performance remains stable. As shown in Table \ref{tab:dacsr:imbtmall}, the performances of ILD metrics are improved by the $p_m(s)$ distribution. This indicates that the modification of distribution with the mask mechanism can mitigate the amplification of major interest.

In conclusion, the relationship between calibration and diversity varies from person to person. For users with homogeneous interests, our model amplifies the effect of the imbalanced distribution. By applying the normalized preference distribution, our model increases the diversity and mitigates the imbalance problem. We proposed a way to provide diversified recommendations for these users under the constraint of calibration. However, whether it is necessary to provide these users with diversified recommendations is a question worth investigating.

\begin{table}[t]
  \centering
  \caption{The performances of distribution $p_m(s)$ on the Tmall dataset}
    \begin{tabular}{c|cccc}
    \toprule
    Models & ILD@10 & $C_{KL}$@10 & ILD@20 & $C_{KL}$@20 \\
    \midrule
    $p(s)$ & 0.6714 & \textbf{2.0114} & 0.7045 & \textbf{1.6240} \\
    $p_m(s)$ & \textbf{0.7201} & 2.0539 & \textbf{0.7516} & 1.6288 \\
    \bottomrule
    \end{tabular}%
  \label{tab:dacsr:imbtmall}%
\end{table}%
\section{Discussion}\label{sec:discussion}

In this paper, we proposed the DACSR model to provide accurate and calibrated recommendation lists for end-to-end sequential recommendation. We conducted experiments on benchmark datasets to demonstrate the effectiveness of our model. 

In general, our model achieved higher accuracy in predicting the next item and provided more calibrated recommendations compared to the post-processing-based model. This is because our model considered the relationship between calibration and accuracy, which was isolated in post-processing-based models. Meanwhile, the end-to-end framework required much less time to provide recommendations than the post-process-based models. 

We displayed the trend in the model's performance as the two main hyperparameters change. As the parameter $\lambda$ varies which stands for the importance of calibration in the objective function, our model achieved better performance in terms of accuracy and calibration. We also analyzed the influence of the parameter $\tau$ which can change the predicted score distribution so that the model focused on items in different score segments.

In the ablation study, we first demonstrated the effectiveness of the proposed loss function for calibration. By applying the calibration loss function, the sequential recommendation models became aware of the preference distribution of recommended items, and aligned it to the historical preference distribution. Therefore, the process of training and prediction was conducted in an end-to-end paradigm. Furthermore, with the decoupled-aggregated framework, the positive information for the calibrated and accurate recommendation was extracted from two individual sequence encoders to improve the performance.

Because calibration is connected to diversity to a certain degree, we finally analyzed the relation between diversity and calibration. We also investigate the effect of imbalanced distribution of homogeneous interests. We found that it differed on the two datasets because of the item attribute coverage and the length of the sequence. For sequences with homogeneous preferences, considering calibration reduced the diversity of recommendations and amplified the main interests of the user.
We adopted two distribution modification methods to improve the diversity and mitigate the effect of imbalanced distribution.

However, there are some limitations in our work:
\begin{itemize}
    \item Our model assigned the objectives of accuracy and calibration to two individual sequence encoders, and combined them with an extraction net to make accurate and calibrated recommendation lists. This structure requires more parameters and time for computation than a single sequence encoder. It would be interesting to explore cleaner models with fewer parameters in the future. 
    \item We focused on providing calibrated recommendations for sequential recommendation models by designing loss functions and the decoupled-aggregated framework. From a different perspective, it would be valuable to investigate the reasons that cause miscalibration in recommendation. For example, previous work has shown there exists correlations between popularity bias and miscalibration \cite{himan20connection}. Besides popularity bias, whether there are other factors contributing to miscalibration and how these factors can be incorporated into the sequence recommendation model are directions worth exploring.
\end{itemize}

\section{Conclusion and Future Work}\label{sec:conclusion}

In this paper, we were committed to exploring the provision of accurate and calibrated recommendation lists based on user behavioural sequences. We proposed a DACSR model to provide accurate and calibrated results for end-to-end sequential recommendation. Specifically, we designed a loss function that estimates the preference distribution of the recommendation list by predicted scores of all items, and measures the consistency with the preference distribution of historical behaviors. In addition, we proposed distribution modification approaches to improve the diversity and mitigate the effect of imbalanced interests. To better handle the goals of accuracy and calibration, we proposed a decoupled-aggregated framework which includes two individual sequence encoders that were assigned with the accuracy and calibration objectives, respectively. Then we utilized an aggregation module to extract information from two sequence encoders to make both accurate and calibrated recommendations. By experiments on benchmark datasets, our model can achieve better accuracy and calibration than the original sequence encoder and the post-processing methods. The ablation studies proved the effectiveness of the general architecture and the extractor net of our model. Finally, we investigated the connection between calibration and diversity, and balanced them by normalizing the historical preference distribution.

For the future work, as mentioned in the discussion part, we first want to explore the cleaner sequence encoder with fewer parameters. Specifically, we may focus on the advanced training paradigm rather than the traditional sequence-to-item paradigm to improve the calibrated sequential recommendation. On one hand, we aim to follow the distillation paradigm, which learns useful information from our DACSR model to make the basic sequence encoder more accurate and calibrated. On the other hand, we also want to explore data augmentation and apply contrastive learning to debias the sequential recommendation models.
In addition, we are also interested in exploring the reasons that cause the miscalibration, and apply them with causal inference to obtain calibrated recommendations. 
Finally, besides the calibrated recommendation, we hope to deal with different types of bias in recommendation algorithms.

\section{Declarations}
The ethical approval is not applicable because our work is not a human or animal study. All authors certify that they have no affiliations with or involvement in any organization or entity with any financial interest or non-financial interest in the subject matter or materials discussed in this manuscript. Author Jiayi Chen contributes to conceptualization, methodology, software, validation, formal analysis, investigation, data curation, writing-original draft, writing-review and editing of the paper. Author Wen Wu contributs to conceptualization, methodology, formal analysis, investigation, writing-Original Draft, writing-review and editing, supervision of the paper. Author Liye Shi contributes to conceptualization, writing-review and editing of this paper. Authors Yu Ji, Wenxin Hu, Xi Chen and Wei Zheng contribute to writing-review and editing of the paper. Author Liang He contributes to writing-review and editing and supervision of the paper. This work is funded by National Natural Science Foundation of China (under project No. 61907016), Science and Technology Commission of Shanghai Municipality, China (under project No. 21511100302), and Natural Science Foundation of Shanghai (under project No. 22ZR1419000). It is also supported by The Research Project of Shanghai Science and Technology Commission (20dz2260300) and The Fundamental Research Funds for the Central Universities. The datasets used in the paper are accessible by the links attached in the content.


\end{document}